\documentclass[%
 reprint,
 amsmath,amssymb,
 aps,
]{revtex4-2}

\usepackage{graphicx}
\usepackage{dcolumn}
\usepackage{bm}
\usepackage{hyperref}
\usepackage{slashed}
\usepackage{subfloat}
\usepackage{pgfplots,subfigure}

\usepackage{geometry}
\definecolor{acsblue}{RGB}{17,76,139}

\begin{document}
\fontsize{7.8}{8.8}\selectfont
\preprint{APS/123-QED}

\title{Photon Rings in Three-Dimensional Bonnor-Melvin Magnetic Spacetime}

\author{Abdullah Guvendi}
\email{abdullah.guvendi@erzurum.edu.tr }
\affiliation{Department of Basic Sciences, Erzurum Technical University, 25050, Erzurum, Türkiye}

\author{Omar Mustafa}
\email{omar.mustafa@emu.edu.tr (Corr. Author)}
\affiliation{Department of Physics, Eastern Mediterranean University, 99628, G. Magusa, north Cyprus, Mersin 10 - Türkiye}

\date{\today}

\begin{abstract}
\vspace{0.15cm}
\setlength{\parindent}{0pt}

{\fontsize{7.8}{8.8}\selectfont We present a rigorous analysis of the relativistic dynamics of vector bosons propagating in a $(2+1)$-dimensional Bonnor-Melvin magnetic spacetime, characterized by an out-of-plane aligned magnetic field and a nonzero cosmological constant $\Lambda$. To achieve this, we derive the exact solution of the fully covariant vector boson equation corresponding to the spin-1 sector of the Duffin-Kemmer-Petiau equation in \(2+1\) dimensions. The magnetic background arises naturally as the $2+1+0$-brane configuration within the Bonnor-Melvin solution to gravity coupled with nonlinear electrodynamics. By deriving the radial wave equation that is mathematically equivalent to the Helmholtz equation governing massless vector bosons, we obtain exact eigenvalue spectra applicable to both massive and massless vector bosons (photons). Remarkably, our results reveal that photons exhibit a finite, nonzero ground-state energy, with their quantum states manifesting solely as rotating ring-like modes. }

\end{abstract}


\maketitle


\section{Introduction}

\vspace{0.15cm}
\setlength{\parindent}{0pt}

Vector bosons, which mediate fundamental interactions in quantum field theory, can be classified as either massive or massless depending on their role in gauge interactions \cite{Griffiths}. Massive vector bosons, such as the \(W^\pm\) and \(Z^0\) bosons of the electroweak interaction, acquire mass via the Higgs mechanism, which spontaneously breaks the \(SU(2)_L \times U(1)_Y\) gauge symmetry \cite{Griffiths,peskin}. Their nonzero mass limits their range, leading to the short-range nature of weak interactions, and introduces a longitudinal polarization state in addition to the two transverse modes, as described by the Proca equation. In contrast, massless vector bosons, exemplified by the photon, mediate long-range interactions, such as electromagnetism, which is governed by an unbroken \(U(1)\) gauge symmetry. Massless vector bosons possess only two transverse polarization states due to gauge invariance and are described by Maxwell's equations, which, in the frequency domain, reduce to the Helmholtz equation \cite{jackson}. This equation governs the behavior of free electromagnetic waves and emerges as the massless limit of the Proca equation, ensuring the preservation of gauge invariance. Investigating the relativistic dynamics of vector bosons in curved spacetime requires a well-established model of their behavior under the influence of gravity \cite{vb-1,vb-2}, with a probabilistic density interpretation. In general relativity, gravity manifests as the curvature of space-time, dictated by the distribution of matter and energy \cite{thorne}. This curvature profoundly influences quantum fields, particularly in regions dominated by strong gravitational and electromagnetic fields \cite{1,2,3,4,5,6,wormhole,epjc,7,8,9,9.1,PM}. A comprehensive understanding of these interactions necessitates the formulation of well-defined, fully covariant vector boson equations \cite{vb-3,cavit}. Furthermore, deriving exact solutions for vector bosons in curved space-time is crucial for advancing our understanding of fundamental interactions across both microscopic and macroscopic scales. 

\vspace{0.15cm}
\setlength{\parindent}{0pt}

On the other hand, magnetic fields, which permeate the universe, play a pivotal role in shaping quantum fields and astrophysical structures, spanning planetary, stellar, and galactic scales. Their influence extends beyond astrophysical phenomena, affecting gravitational systems and high-energy processes, thus making magnetized space-time models fundamental in theoretical physics. A notable example is the Bonnor-Melvin magnetic (BMM) space-time, which constitutes an exact solution to Einstein’s field equations \cite{r7,r8,r9,r10,r100,r200}. A generalization of this model that incorporates a nonzero cosmological constant \(\Lambda\) extends its applicability by modifying the space-time curvature and influencing both local and global gravitational phenomena \cite{r10}. This extension provides a broader framework for understanding gravitational and high-energy astrophysical processes within the paradigm of general relativity \cite{r7,r8,r9,r10,r100,r200}. The impact of such a magnetized space-time on quantum fields has been a subject of considerable interest, with studies exploring its effects on diverse quantum fields, including non-interacting and interacting Dirac and Weyl fermions \cite{10,11,11-b}, scalar bosons \cite{12,13,14,15,16} and fermionic fields \cite{17,18}. The cylindrical symmetry inherent in this space-time plays a crucial role in simplifying the mathematical treatment of quantum fields, as it ensures invariance under Lorentz boosts along the symmetry axis \cite{19,20}. Consequently, the behavior of relativistic quantum systems in such backgrounds adheres to both special relativity and quantum mechanics, imposing strict transformation properties on the corresponding field equations. The dynamical symmetries of these systems are of fundamental significance not only in high-energy physics and cosmology but also in condensed matter physics, where they provide insights into emergent phenomena \cite{21,22,23,24,25,26}. 

\vspace{0.15cm}
\setlength{\parindent}{0pt}

In this study, we investigate relativistic vector bosons in a curved (2+1)-dimensional magnetic spacetime with a nonzero cosmological constant, \(\Lambda\). This background generalizes the original four-dimensional BM solution \cite{r10,r100,r200} by incorporating the effects of \(\Lambda\), thus modifying the spacetime curvature and influencing the eigenvalue spectra of vector bosons in a nontrivial manner. Our objective is to derive exact solutions for vector bosons in the three-dimensional generalized magnetic background, which corresponds to the \((2+1+0)\)-brane solution of the BMM spacetime in gravity coupled to nonlinear electrodynamics \cite{r200}, and to explore the impact of topological parameters on their dynamical properties. The results provide new insights into the fundamental interactions of vector bosons in curved spacetime and their behavior in a magnetized cosmological setting. This manuscript is organized as follows: Section \ref{sec:2} presents the theoretical framework and mathematical formulation. In Section \ref{sec:3}, we derive exact solutions for vector bosons in a curved background with a cosmological constant and analyze their eigenvalue spectra. Finally, Section \ref{sec:4} summarizes our key findings and discusses their implications for various physical scenarios. Throughout this work, we employ natural units, where \(c=1=\hbar=4\pi G\).

{\section{Bonnor-Melvin geometric confinement}} \label{sec:2}

In this section, we first introduce the three-dimensional BMM space-time background, which plays a crucial role in understanding the quantum dynamics of fields in curved and magnetized geometries. This space-time is described by the following metric \cite{r10, r100, r200}, employing the $(+, -, -)$ signature:
\begin{equation}
ds^2 = g_{\mu\nu} \, dx^\mu \, dx^\nu = dt^2 - d\rho^2 - \eta^2(\rho) \, d\phi^2.\label{metric}
\end{equation}
Here, the function \(\eta(\rho)\) is defined as \(\eta(\rho) = \sigma \sin \left( \sqrt{2\Lambda} \rho \right)\), where \(\Lambda > 0\). In this expression, \(\sigma\) represents an integration constant, and the cosmological constant \(\Lambda\) serves as a fine-tuning parameter for the magnetic field strength. The magnetic field is given by \(H = \sqrt{\Lambda}\, \sigma \sin \left( \sqrt{2\Lambda} \rho \right)\), as detailed in \cite{r10, r100, r200}. The Greek indices \(\mu, \nu = t, \rho, \phi\) correspond to the coordinates of the curved space-time. The focus here is on the quantum dynamics of particles residing on a two-dimensional curved surface \(M\). For a compact surface, the Gauss-Bonnet theorem provides an essential result, stating that the integral of the Gaussian curvature \(K\) over the surface \(M\) is equal to \(2\pi\) times the Euler characteristic \(\chi\) of \(M\), i.e., \(\int_M K \, dA = 2\pi \chi\). In \(2+1\) dimensions, the Euler characteristic \(\chi\) of a closed surface is \(2\), leading to the result \(\int_M K \, dA = 4\pi\). The Ricci scalar \(R\) measures the intrinsic curvature of the space and, in \(2+1\) dimensions, is proportional to the integral of the Gaussian curvature \(K\). Hence, we have \(R = 2K\), as the integral of \(K\) over the surface equals \(4\pi\), as described in \cite{18}. For the metric described by equation \eqref{metric}, the covariant metric tensor is \(g_{\mu\nu} = \text{diag}(1, -1, -\eta^2(\rho))\), and its inverse is given by: \(g^{\mu\nu} = \text{diag}(1, -1, -\eta^{-2}(\rho))\). The determinant of the metric tensor is \(g = \det{g_{\mu\nu}} = \eta^2(\rho)\). In this framework, the non-zero components of the Christoffel symbols \( \Gamma^{\mu}_{\nu \zeta} \) can be derived and are found to be:
\(\Gamma^{\rho}_{\phi\phi} = -\eta \, \eta_{,\rho}, \quad \Gamma^{\phi}_{\rho\phi} = \Gamma^{\phi}_{\phi \rho} = \frac{\eta_{,\rho}}{\eta}\) \cite{18}. The comma denotes a derivative with respect to the coordinate \( \rho \). For the metric \eqref{metric}, the only non-vanishing component of the curvature tensor is \(R_{\rho\phi \rho\phi} = \eta \, \eta_{,\rho\rho}\). As a result, the Ricci tensor \( R_{\mu\nu} \) has the non-zero components: \(R_{\rho\rho} = \frac{\eta_{,\rho\rho}}{\eta}, \quad R_{\phi\phi} = \eta \, \eta_{,\rho\rho}\). The scalar curvature, or Ricci scalar \( R \), is obtained by contracting the Ricci tensor with the inverse metric: \(R = g^{\mu\nu} \, R_{\mu\nu} = -2 \, \frac{\eta_{,\rho\rho}}{\eta}\) \cite{18}. Therefore, the Gaussian curvature \(K\) of this space-time background is given by:
\begin{equation}
K = -\frac{\eta_{,\rho\rho}}{\eta} = 2 \Lambda, \quad \text{where} \quad \Lambda > 0.
\end{equation}
It is crucial to highlight that the singular behavior of the function \(\eta(\rho)\) and the magnetic field \(H\), resulting from their oscillatory nature, plays a central role in the structure of spacetime. Specifically, \(\eta(\rho)\) vanishes at discrete points \(\rho = \frac{\kappa\pi}{\sqrt{2\Lambda}}\), where \(\kappa \in \mathbb{Z}_{\geq 0}\) (i.e., \(\kappa\) is a non-negative integer). At these points, the metric component \(g_{\phi\phi}\) degenerates, leading to a breakdown in the spacetime structure. This degeneration signals the appearance of a singularity in the spacetime, often associated with topologically stable defects, such as effective confinement walls. Such singularities in \(\eta(\rho)\) mark regions where field configurations undergo sharp transitions, with potential implications for the formation of matter or photon rings within the spatial domain \(\rho \in \left[0, \frac{\pi}{\sqrt{2\Lambda}}\right]\). The degeneration of \(g_{\phi\phi}\) at these specific points creates critical locations in spacetime where the smooth geometry is disrupted, facilitating the emergence of localized phenomena and topological features in the system. These features suggest a direct connection between the geometry of spacetime and the electromagnetic field. The vanishing of \(\eta(\rho)\) at certain \(\rho\)-values leads to undefined or degenerate metric components, highlighting the presence of localized defects or boundaries. This, in turn, indicates the formation of topological defects and electromagnetic confinement, all of which are tightly linked to the singular behavior of the metric within the specified spatial region. This property demands rigorous investigation, as it has profound implications for the understanding of spacetime structure and the dynamics of electromagnetic fields, particularly in the context of localized phenomena and field confinement within specific spatial domains. It is important to clarify that the surfaces at 
\(\rho = \frac{\pi}{\sqrt{2\Lambda}}\) are not domain walls in the strict field-theoretic sense, where a domain wall separates distinct vacuum phases of a field. Instead, these surfaces correspond to geometric or topological defects arising from the degeneration of the angular metric component \(g_{\phi\phi}\). At these radii, the compact \(\phi\)-dimension collapses, producing singularities in the spacetime structure. Consequently, these boundaries act as effective confinement walls for fields within the radial domain, similar to domain walls phenomenologically, but their origin is purely geometric rather than associated with a phase transition of the field. This distinction emphasizes the role of topology and geometry in the confinement mechanism present in the BMM spacetime.\\

\section{Photonic modes} \label{sec:3}

In this section, we first write the generalized fully covariant vector boson equation corresponding to the spin-1 sector of the Duffin-Kemmer-Petiau formalism in \(2+1\) dimensions \cite{vb-3,cavit} and subsequently specialize it to describe relativistic vector bosons propagating in the specified three-dimensional magnetized background with a nonvanishing cosmological constant \(\Lambda\). We derive a coupled system of three equations, one of which is purely algebraic, and then proceed to obtain exact solutions for this system. In \(2+1\)-dimensions, the generalized vector boson equation takes the form \cite{vb-1, vb-2}:
\begin{eqnarray}
\left\lbrace \mathcal{B}^{\mu}\slashed{\nabla}_{\mu}+im_{b}\textbf{I}_{4}\right\rbrace \Psi\left(x^{\mu}\right)=0, \label{eq-7}
\end{eqnarray}
where \( \slashed{\nabla}_{\mu} \) denotes the covariant derivatives, defined by \( \slashed{\nabla}_{\mu}=\partial_{\mu}-\Omega_{\mu} \). The matrices \( \mathcal{B}^{\mu} \) are position-dependent and constructed from the generalized Dirac matrices \( \gamma^{\mu} \), given by \( \mathcal{B}^{\mu}=\frac{1}{2}\left[\gamma^{\mu}\otimes \textbf{I}_{2}+\textbf{I}_{2}\otimes\gamma^{\mu}\right] \). Here, \( m_{b} \) represents the rest mass of the vector boson, and \( \Psi\left(x^{\mu}\right) \) is the symmetric spinor that depends on space-time coordinates. Additionally, \( \textbf{I}_{4} \) and \( \textbf{I}_{2} \) are the four- and two-dimensional identity matrices, respectively. The symbol \( \otimes \) denotes the Kronecker product, while \( x^{\mu} \) refers to the space-time position vector. The affine spin connections \( \Omega_{\mu} \) for vector fields are determined from the spinorial affine connections \( \Gamma_{\mu} \) for Dirac fields, as given by \( \Omega_{\mu}=\Gamma_{\mu}\otimes \textbf{I}_{2}+\textbf{I}_{2}\otimes \Gamma_{\mu} \) \cite{vb-1, vb-2}. The generalized Dirac matrices \( \gamma^{\mu} \) and the nonzero components of the affine spin connections \( \Gamma_{\mu} \) for Dirac fields are provided in \cite{10,19}:
\begin{equation}
\gamma^t = \sigma^z, \quad \gamma^\rho = i \sigma^x, \quad \gamma^\phi = i \eta^{-1} \sigma^y,\label{DM}
\end{equation}
and
\begin{equation}
\Gamma_\phi = \frac{i}{2} \eta_{,\rho} \sigma^z,\label{AC}
\end{equation}
where \( \sigma^x, \sigma^y, \sigma^z \) are the Pauli matrices, and \( i = \sqrt{-1} \). The space-time-dependent vector boson field \( \Psi\left(x^{\mu}\right) \) is constructed as the direct product of two symmetric Dirac spinors \cite{vb-1,vb-2}, and it can be factorized as follows:
\begin{eqnarray}
 \Psi\left(x^{\mu}\right)=\textrm{e}^{-i\,\mathcal{E}\, t}\,\textrm{e}^{i\,s\, \phi}\, \left(\psi_{1}(\rho) ,\,\psi_{2}(\rho),\,\psi_{3}(\rho),\,\psi_{4}(\rho)\right)^{T},\label{eq-8}
\end{eqnarray}
where \( \mathcal{E} \) denotes the relativistic energy, and \( s \) represents the spin of the vector boson. By substituting Eq. \eqref{DM}, Eq. \eqref{AC} and Eq. \eqref{eq-8} into Eq. \eqref{eq-7}, we derive a matrix equation, leading to four first-order differential equations. By adding and subtracting these equations, we obtain the following system (see also \cite{vb-1,vb-2}):
\begin{eqnarray}
&\mathcal{E}\,\psi_{+} -m_{b}\, \psi_{-} -\frac{s}{\eta}\psi_{0}=0,\quad \mathcal{E}\,\psi_{-} -m_{b}\, \psi_{+} -\psi_{0_{,\rho}}=0,\nonumber\\
&m_{b}\,\psi_{0} -\frac{s}{\eta}\,\psi_{-}+ \hat{\lambda}\,\psi_{+}=0,\label{eq-9}
\end{eqnarray}
where \( \psi_{0}=\psi_{2}+\psi_{3} \), \( \psi_{\pm}=\psi_{1}\pm\psi_{4} \), and \( \hat{\lambda}=\partial_{\rho}+\frac{\eta_{,\rho}}{\eta} \). Solving this system for \( \psi_{0} \) yields the following equation:
\begin{eqnarray}
\psi_{0_{,\rho\rho}}+ \frac{\eta_{,\rho}}{\eta} \psi_{0_{,\rho}}+\left(\epsilon^2-\frac{s^2}{\eta^2}\right)\,\psi_{0}=0,
\end{eqnarray}
where \( \epsilon^2=\mathcal{E}^2-m_{b}^2 \). Here, it is important to emphasize that this wave equation corresponds to the Helmholtz equation when \(m_b^2 = 0\), with the substitution of \(\mathcal{E}\) by the wave number. In this form, it describes the dynamics of photons without any loss of generality. Next, we proceed to derive the effective potential governing vector bosons in the BMM background. By introducing the variable transformation \( z = \rho \sqrt{2\Lambda} \), we arrive at:
\begin{equation}
\left( \partial^2_{z}+\cot(z)  \partial_{z} -\frac{\tilde{s}^2}{\sin(z)^2}+\tilde{\epsilon}^2\right)\,\psi_{0}(z) = 0,
\end{equation}
where \( \tilde{s}=\frac{s}{\sigma\sqrt{2\Lambda}} \) and \( \tilde{\epsilon}=\frac{\epsilon}{\sqrt{2\Lambda}} \). Eliminating the first-order derivative via the transformation:
\(\psi_{0}(z) = \sin\left(z\right)^{-1/2} \psi(z)\) yields the Schr\"odinger-like equation:
\begin{equation}
\begin{split}
&\psi''(z)+\left(\tilde{\mathcal{E}}^2-V_{eff}(z)\right)\psi(z)=0,\quad \tilde{\mathcal{E}}=\frac{\mathcal{E}}{\sqrt{2\Lambda}},\\
&V_{eff}(z)=\tilde{m}_{b}^2+\frac{(\tilde{s}^2-1/4)}{\sin(z)^2}-\frac{1}{4},\quad \tilde{m}_{b}=\frac{m_{b}}{\sqrt{2\Lambda}}.
\end{split} \label{Veff}
\end{equation}
This effective potential exhibits significant structural and asymptotic characteristics due to its singular and periodic nature. The term \( 1/\sin^2(z) \) introduces singularities at \( z = \kappa\pi \), where the potential diverges. If \( \tilde{s}^2 > \frac{1}{4} \), these singularities form impenetrable barriers, whereas for \( \tilde{s}^2 < \frac{1}{4} \), they create attractive wells capable of supporting localized states. Near \( z = 0 \), using \( \sin(z) \approx z \), the potential behaves as \( V_{\text{eff}}(z) \approx \tilde{m}_b^2 + \frac{\tilde{s}^2 - \frac{1}{4}}{z^2} - \frac{1}{4} \), exhibiting an inverse-square term influencing wave function behavior. The structure of this potential governs spatial constraints, and energy quantization, emphasizing its fundamental role in determining the vector boson dynamics. Now, let us determine the wavefunction and energy eigenvalues in a closed form. To achieve this, we proceed with a second change of variable, \(x = \sin(z)\), which transforms the equation into
\begin{equation}
\left( x^{2}-1\right) \psi^{\prime \prime }\left( x\right) +x\psi^{\prime }\left(
x\right) +\left( \frac{\tilde{s}^{2}-1/4}{x^{2}}-\tilde{\varrho}\right)
\psi\left( x\right) =0,  
\end{equation}
where \(\tilde{\varrho}=\tilde{\epsilon}^2+1/4\). Introducing yet another change of variable, \(u = x^{2}\), we obtain
\begin{equation}
\left( u^{2}-u\right) \psi^{\prime \prime }\left( u\right) +\left( u-\frac{1}{2}
\right) \psi^{\prime }\left( u\right) +\left( \frac{\zeta}{u}-\frac{\tilde{
\varrho}}{4}\right) \psi\left( u\right) =0,  
\end{equation}
where \(\zeta=\frac{\tilde{s}^{2}}{4}-\frac{1}{16}\). At this stage, it is evident that a power series solution can be employed in the form of \(\psi\left( u\right) =\sum\limits_{j=0}^{\infty }B_{j}\,u^{j+k }\). Substituting this into the transformed equation leads to
\begin{equation}
\begin{split}
&\sum\limits_{j=0}^{\infty }B_{j}\left[ \left( j+k \right) ^{2}-\frac{\tilde{\varrho}}{4}\right] \,u^{j+k }+B_{0}\left[ \zeta-k \left( k -\frac{1}{2}\right) \right] \,u^{k -1}\\
&+\sum\limits_{j=0}^{\infty }B_{j+1}\left[ \zeta-\left( j+k +1\right) \left( j+k +\frac{1}{2}\right) \right] \,u^{j+k }=0.  
\end{split}
\end{equation}
Since \(B_{0}\neq 0\), it follows that \(\zeta-k \left( k -\frac{1}{2}\right) =0\Longrightarrow k =\frac{1}{4}\pm \frac{\left\vert \tilde{s}\right\vert }{2}\). To ensure that the radial function satisfies \(\psi\left( u\right) \longrightarrow 0\) as \(u\longrightarrow 0\), we adopt the solution \(k =\frac{1}{4}+\frac{\left\vert \tilde{s}\right\vert }{2}\), considering the allowed range \(z\in \left[0,\pi \right]\).
Consequently, we derive the two-term recurrence relation,
\begin{gather}
B_{j}\left[ \left( j+k \right) ^{2}-\frac{\tilde{\varrho}}{4}\right] +B_{j+1}\left[ \zeta-\left( j+k +1\right) \left( j+k +\frac{1}{2}\right) \right]  =0,  
\end{gather}
establishing a correlation between the power series coefficients. By truncating the series to a polynomial of order \(n\geq 0\), the condition \(B_{n+1}=0\) with \(B_{n}\neq 0\) implies that \(\tilde{\varrho}=\left( 2n+\left\vert \tilde{s}\right\vert +\frac{1}{2}\right) ^{2} \) where $s=\pm1,\pm2,..\,$. Thus, the wave functions take the form \(\psi\left( u\right) =\mathcal{C}\,
u^{\left\vert \tilde{s}\right\vert /2+1/4}\sum\limits_{j=0}^{n}B_{j}\,u^{j}\), which, in terms of the original variable \(\rho\), transforms into \(\psi\left( \rho\right) =\mathcal{C}\,\sin \left( \rho\,\sqrt{2\Lambda}\right) ^{\left\vert \tilde{s}\right\vert+1/2}\sum\limits_{j=0}^{n}B_{j}\,u^{j}\). Consequently, the radial wavefunction is given by
\begin{figure}[ht!]
   \centering
   \includegraphics[scale=0.42]{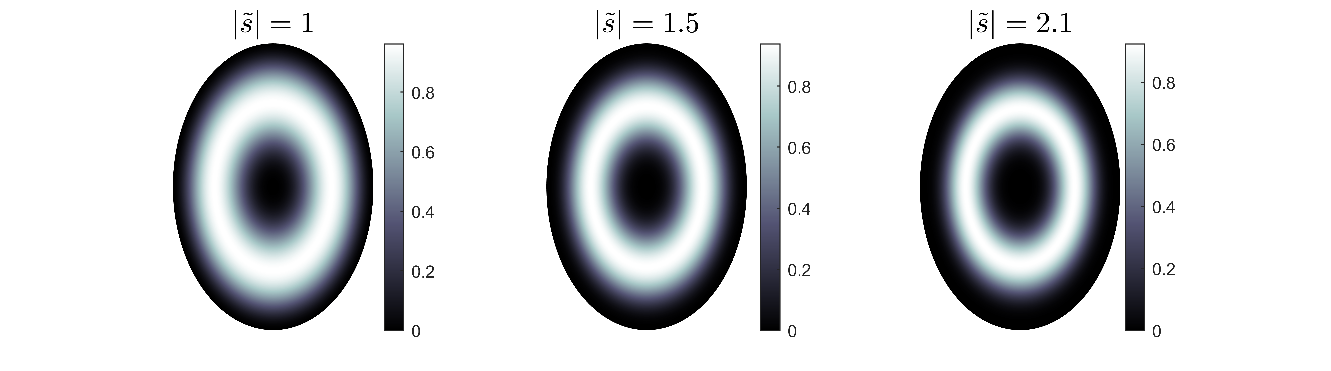}\\
   \caption{ \footnotesize Plots of the radial probability density function in two dimensions (\(X, Y\)), where \(X = z\, \cos(\phi)\) and \(Y = z\, \sin(\phi)\). The three subplots correspond to different values of \(\tilde{s} = 1, 1.5, 2.1\), demonstrating how the probability distribution changes with increasing \(\tilde{s}\).}\label{fig:1}
\end{figure}
\begin{equation}
\psi_{0_{n,s}} \left( \rho\right) =\mathcal{C}\,\sin \left( \rho\,\sqrt{2\Lambda}\right) ^{\left\vert \tilde{s}\right\vert}\sum\limits_{j=0}^{n}B_{j}\,u^{j}, \label{WF} 
\end{equation}
where \(u=x^{2}=\sin \left( \rho\,\sqrt{2\Lambda}\right) ^{2}\). It is noteworthy that the polynomial in this expression corresponds to a hypergeometric polynomial with even powers of \(\sin(z)\). Furthermore, the function satisfies the boundary conditions \(\psi(0)=0=\psi(\pi)\), as expected for a textbook infinite potential well. From the above derivation, the energy spectrum follows as
\begin{equation}
\mathcal{E}_{n,s}=\pm m_{b}\sqrt{1+\frac{2\Lambda}{m^{2}_{b}} \left( 2n+\left\vert \tilde{s}\right\vert \right) \left( 2n+\left\vert \tilde{s}\right\vert +1\right)}.\label{ES-m}
\end{equation}
The results suggest that the parameter \(\Lambda\) governs the relativistic oscillator-like behavior of the bosons, as described in \cite{23}. In our analysis, the intrinsic spin of the vector bosons becomes topologically modified, since \( \tilde{s} = \frac{s}{\sigma \sqrt{2\Lambda}} \), acquiring a nontrivial dependence on both the cosmological constant and the topological parameter \(\sigma\). This modification profoundly affects the spin-statistics correspondence, such that the resulting excitations are no longer restricted to conventional bosonic behavior. Instead, the altered spin structure opens the possibility for the emergence of anyonic statistics-a fractional form of quantum statistics typically realized only in two-dimensional systems. Remarkably, this implies that photons in the BMM spacetime may exhibit anyonic properties, reflecting how nontrivial spacetime topology and magnetization can induce fundamentally exotic quantum states. Such behavior closely aligns with foundational results on fractional statistics and topological phases in lower-dimensional quantum field theories \cite{wilczek1982,catenoid}. Furthermore, the expression in \eqref{ES-m} holds true for massless vector bosons when \(m^2_{b}=0\), which consequently leads to the photonic ground state(s) for \(n=0\)
\begin{equation}
\mathcal{E}_{0,s}=\pm \sqrt{2\Lambda \left(\tilde{s}^2 +\left\vert \tilde{s}\right\vert\right)}\Rightarrow \mathcal{E}_{0,0}=0.
\end{equation}
Now, let us return to the wave functions given by \eqref{WF}. In terms of the variable \( z \), we can express the ground state wavefunction(s) for photons, and accordingly, we can determine the corresponding radial probability density function(s) as follows:
\begin{equation}
P_{0,s}=\left\vert \mathcal{C}\right\vert^2 \, \int \sin \left( z\right) ^{2\,\left\vert \tilde{s}\right\vert}\,z\,dz.\label{RPDF}
\end{equation}
It is important to note that this result is valid under the conditions \(\Lambda \neq 0\) and \(\sigma \neq 0\), with the relations \(z = \rho \sqrt{2\Lambda}\) and \(\tilde{s} = \frac{s}{\sigma \sqrt{2\Lambda}}\). The corresponding ground-state wavefunction and energy eigenvalues suggest that the modes are associated with spin in the magnetic background. These states can be interpreted as rotating ($s\neq 0$) ring-like modes or, equivalently, as rotating magnetic vortices, regardless of whether \(m_b = 0\). This behavior is illustrated in Figure \ref{fig:1}. As the parameter $|\tilde{s}|$ increases, these peaks become sharper and more pronounced, resulting in more localized and narrowly confined probability distributions around specific radii. Physically, this reflects states with higher quantum numbers having tighter radial confinement, with the particle or excitation being most likely found near distinct radial distances. Furthermore, our results demonstrate that the magnetized spacetime background is pivotal for the formation of photon rings, as it permits angular propagation within a finite radial confinement region.

\vspace{0.2cm}
\setlength{\parindent}{0pt}

On the other hand, in the limit of small \(\Lambda\), we approximate the sine function as \(\sin(\sqrt{2\Lambda} \, \rho) \approx \sqrt{2\Lambda} \, \rho\), which leads to the following spacetime metric:
\[
ds^2 \approx dt^2 - d\rho^2 - \alpha^2 \rho^2 \, d\phi^2, \quad \alpha = \sigma\sqrt{2\Lambda}.
\]
The radial dependence of \(\eta(\rho)\) gives rise to a conical geometry, similar to the spacetime around static cosmic strings or point-like defects, which are characterized by deficit angles (\(\alpha \in (0,1]\)) \cite{vilenkin,katanaev}. The inclusion of a magnetic field modifies the angular part of the metric, creating a defect-like structure that alters both the topology and symmetry of the spacetime. This alteration introduces strong local curvature, especially at small radial distances. Moreover, by making the identification \(\phi \rightarrow -\frac{i}{\tilde{\alpha}\ell}\,t\), the metric transforms into:
\[
ds^2 = \frac{\tilde{\alpha}^2 \rho^2}{\ell^2} dt^2 - d\rho^2 - \tilde{\alpha}^2 \ell^2 \, d\phi^2,
\]
where \(\tilde{\alpha}^2=\alpha\) (with \(0 < \tilde{\alpha}^2  \leq 1\)) is related to the black hole mass, and \(\ell\) is associated with the cosmological constant \cite{corichi}. As discussed earlier, this form describes the near-horizon region of the static BTZ black hole \cite{vb-2,corichi}. As a result, the BMM spacetime provides a versatile framework for investigating cosmological phenomena \cite{20}, condensed matter systems \cite{hollow}, and quantum criticality \cite{criticality,criticality-2}.

\section{Summary and discussions}\label{sec:4}

In this research, we investigated the influence of the BMM spacetime on the relativistic dynamics of both massive and massless vector bosons. This cylindrically symmetric, magnetized spacetime is characterized by a homogeneous magnetic field aligned along the symmetry axis and a nonzero cosmological constant \( \Lambda \). The background preserves Lorentz invariance under boosts in the \( z \)-direction, which allows us to examine the dynamics of vector bosons within its \((2+1)\)-dimensional formulation. This structure is derived as the BMM \((2+1+0)\)-brane solution in gravity coupled to nonlinear electrodynamics. The three-dimensional magnetic background is mathematically described by the metric: \(ds^2 = g_{\mu\nu} \, dx^\mu \, dx^\nu = dt^2 - d\rho^2 - \eta^2(\rho) \, d\phi^2\), where \(\eta(\rho)\) is defined as: \(\eta(\rho) = \sigma \sin \left( \sqrt{2\Lambda} \rho \right)\). By analyzing the dynamics of vector bosons in this background, we have shown that the effective gravitational potential \(V_{\text{eff}}(\rho)\), emerging naturally from the Bonnor--Melvin spacetime geometry, acts as an infinitely high confining barrier. These boundaries at \(\rho \in \left[0, \frac{\pi}{\sqrt{2\Lambda}}\right]\) arise from the degeneration of the angular metric component and are better interpreted as geometric or topological defects, which effectively confine the particle motion within this radial domain.

\vspace{0.15cm}
\setlength{\parindent}{0pt}

Focusing on this intriguing property, we derive the wave equation that governs the motion of vector bosons. Remarkably, in the massless limit \(m_b^2 = 0\), this equation reduces to the standard Helmholtz equation. This result indicates that our model is directly applicable to electromagnetic waves without loss of generality, allowing us to extend our analysis to the behavior of photons in this background. It is evident that massive and massless vector bosons are constrained to remain indefinitely within the aforementioned radial confinement region. A crucial outcome of our investigation is that the corresponding one-dimensional Schrödinger-like equation admits exact analytical solutions. The wave functions obtained take the form of standard polynomials, as given in Eq. (\ref{WF}), satisfying the necessary conditions of finiteness in \(\rho=0\) and \(\rho=\frac{\pi}{\sqrt{2\Lambda}}\), as well as square integrability in the allowed domain. For photons, we derive exact energy eigenvalues and eigenfunctions. In particular, for ground-state photons, the energy spectrum is obtained as: \(\mathcal{E}_{0,s} = \pm \sqrt{2\Lambda \left(\tilde{s}^2 + \left\vert \tilde{s} \right\vert \right)}\), where \(\tilde{s} = \frac{s}{\sigma\sqrt{2\Lambda}}\), with the corresponding radial wave function: \(\psi_{0_{0,s}}(\rho) \propto \sin (\rho \sqrt{2\Lambda})^{\left\vert \tilde{s} \right\vert}\). Another key feature revealed by our analysis is the emergence of a topologically modulated spin parameter, \(\left\vert \tilde{s} \right\vert\), which depends on the topology of the spacetime or, equivalently, on the magnitude of the background magnetic field. This observation suggests that variations in the topological structure or magnetic field strength lead to modifications in the photonic spin states. Moreover, our results demonstrate that photons acquire nonzero ground-state energies for different \(\tilde{s}\) states at \(n=0\), highlighting an intrinsic energy shift induced by the background topology. Furthermore, our solutions provide the radial probability function, which describes the probability distribution of photonic modes. We establish that these states manifest as rotating ring-like photonic modes, a direct consequence of the fact that \(\tilde{s} \neq 0\) if and only if \(\Lambda \neq 0\) and \(\sigma \neq 0\). This observation reinforces the notion that the background structure inherently supports the formation of ring-like photonic modes (see Figure \ref{fig:1}). The magnetic field in this background is given by the function: \(H(\rho) = \sqrt{\Lambda}\, \sigma\, \sin (\sqrt{2\Lambda} \rho)\), as detailed in prior studies \cite{r10, r100, r200}. Consequently, the rotating, ring-like photonic modes can be viewed as topologically protected spin states, since only configurations with nonvanishing spin \((|\tilde{s}| \neq 0)\) are permitted, and they remain confined within the magnetized background. At this juncture, it is important to emphasize that, while the radial motion is strongly constrained by the background geometry, the angular degree of freedom remains unimpeded. This naturally leads to the formation of photon-ring structures, which may hold significant relevance for both optical analogues and astrophysical phenomena. Our findings, therefore, may open a conceptual pathway for studying confined photonic modes in magnetized curved spacetimes, with prospective applications in fundamental physics and emergent photonic technologies. In particular, the \(2+1\)-dimensional Bonnor Melvin magnetic spacetime, characterized by a radially modulated spatial geometry and a built-in magnetic field, can be meaningfully emulated in laboratory settings via analogue gravity techniques. The metric profile \( \eta(\rho) = \sigma \sin(\sqrt{2\Lambda} \rho) \) can, for instance, be implemented using transformation optics in metamaterials, where spatially engineered anisotropic permittivity and permeability distributions recreate the desired geodesic structure for light propagation~\cite{Leonhardt2009}. Related strategies have been successfully employed to simulate curved backgrounds and celestial-like dynamics in photonic crystals and waveguide arrays~\cite{Genov2009}, offering a feasible route toward realizing Bonnor Melvin type geometries. Furthermore, synthetic magnetic fields with tunable spatial profiles, crucial for reproducing the magnetic sector of this spacetime, have been experimentally generated in ultracold atomic gases using laser-induced gauge potentials~\cite{Dalibard2011} and in Floquet-engineered photonic lattices~\cite{Rechtsman2013}, where temporal modulations induce effective gauge fields for photons. Collectively, these analogue frameworks provide concrete, experimentally accessible models for exploring wave confinement, vortex-like excitations, and anyonic phenomena in effectively curved and magnetized spacetimes, thereby grounding the broader condensed matter and photonic implications of our results.

{\section*{Acknowledgments}
The authors sincerely thank the reviewer for their thoughtful comments and constructive feedback, which have helped to improve the clarity and quality of this manuscript.}

\section*{Credit authorship contribution statement}

\textbf{Abdullah Guvendi}: Conceptualization, Methodology, Formal Analysis, Writing – Original Draft, Investigation, Visualization, Writing – Review and Editing.\\
\textbf{Omar Mustafa}: Conceptualization, Methodology, Formal Analysis,  Writing – Original Draft, Investigation, Writing – Review and Editing. 

\section*{Data availability}

This manuscript has no associated data.

\section*{Conflicts of interest statement}

No conflict of interest declared by the authors.

\section*{Funding}

No fund has received for this study.

\end{document}